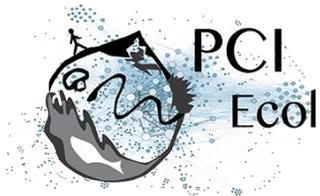

*Title:* The importance of sampling design for unbiased estimation of survival using joint live-recapture and live resight models

Authors: Dzul[1,3], Maria C., Charles B. Yackulic[1], William L. Kendall[2]

[1]U.S. Geological Survey
Southwest Biological Science Center
Grand Canyon Monitoring and Research Center
2255 N. Gemini Dr.
Flagstaff, AZ, 86001, USA

[2]U.S. Geological Survey
Colorado Cooperative Fish and Wildlife Research Unit
Colorado State University
201 J.V.K. Wagar Building 1484 Campus Delivery
Fort Collins, CO 80523, USA

[3]Corresponding author, mdzul@usgs.gov

*Abstract:*

Survival is a key life history parameter that can inform management decisions and basic life history research. Because true survival is often confounded with emigration from the study area, many studies are forced to estimate apparent survival (i.e., probability of surviving and remaining inside the study area), which can be much lower than true survival for highly mobile species. One method for estimating true survival is the Barker joint live-recapture/live-resight (JLRLR) model, which combines capture data from a study area (hereafter the 'capture site') with resighting data from a broader geographic area. This model assumes that live resights occur throughout the entire area where animals can disperse to and this assumption is often not met in practice. Here we use simulation to evaluate survival bias from a JLRLR model under study design scenarios that differ in the site selection for resights: global, random, fixed including the capture site, and fixed excluding the capture site. Simulation results indicate that fixed designs that included the capture site showed negative survival bias, whereas fixed designs that excluded the capture site exhibited positive survival bias. The magnitude of the bias was dependent on movement and survival, where scenarios with high survival and frequent movement had minimal bias. In effort to help minimize bias, we developed a multistate version of the JLRLR and demonstrated reductions in survival bias compared to the single-state version for most designs. Our results suggest minimizing bias can be accomplished by: 1) using a random resight design when feasible and global sampling is not possible, 2) using the multistate JLRLR model when appropriate, 3) including the capture site in the resight sampling frame when possible, and 4) reporting survival as apparent survival if fixed sites are used for resight with the single state JLRLR model.

*Introduction:*

Survival can be an indicator of population health and can inform our understanding of life history strategies in many species (Marshall *et al.* 2004; Gilroy *et al.* 2012). Moreover, survival estimates often have relevance to species management, in particular when managers must set harvest limits (Sedinger & Rexstad 1994, Francis, Sauer & Serie 1998, Post *et al.* 2003) or evaluate conservation actions (Doherty *et al.* 2014, Oppel *et al.* 2016, Yackulic *et al*. 2021). However, for highly mobile species, estimating true survival in field settings is difficult because survival estimates are confounded with permanent emigration from the study area (Seber 1982; Lebreton *et al.* 1992; Schaub *et al.* 2004). As a result, many studies can only measure apparent survival (i.e., the probability an animal is still alive and did not permanently move out of the capture site), often using the Cormack-Jolly-Seber (CJS) model (Cormack 1964; Jolly 1965; Seber 1965). From a manager's perspective, apparent survival estimates are dissatisfying because movement and mortality have very different implications for population persistence and habitat conservation (Ergon & Gardner 2014). For example, if emigration is high, conservation efforts may focus on habitat connectivity and creating migration corridors outside the study area (DeMaynadier & Hunter Jr 1999), whereas if mortality is high, conservation efforts may focus on improving habitat within the study area (Sun *et al.* 2003; Trumbo *et al.* 2021).

While numerous methods have been developed to account for the effects of temporary (Kendall, Nichols & Hines 1997; Ergon & Gardner 2014) and permanent emigration (Oro *et al.* 2004; Gilroy *et al.* 2012; Badia-Boher *et al.* 2023) on estimation of survival, here we focus on differentiating survival from permanent emigration using the Barker joint live-recapture/live-resight (hereafter JLRLR model) model (Barker 1997). Whereas most mark-recapture models require detections to occur in discrete time, the JLRLR model is designed to provide non-biased survival estimates when resight data are continuous over a long interval. Specifically, the JLRLR model pairs discrete capture events that occur in a localized study area (capture site) with long resight intervals from a broader geographic region. One additional benefit to the JLRLR model is that, since it accounts for movement in and out of a capture site, it can be used to estimate true survival in situations where animals emigrate permanently from the capture site. Importantly, however, information about the location of resight (e.g., inside or outside the study area) is not included in the JLRLR model.

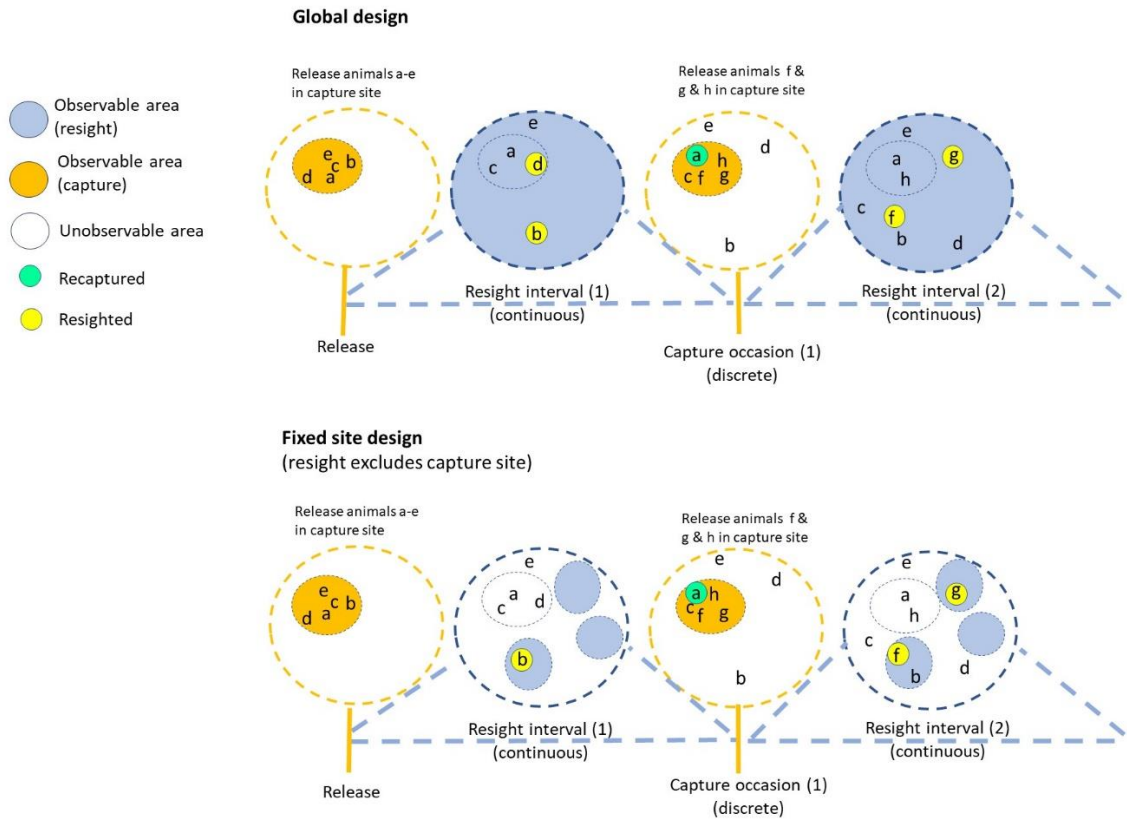

*Figure 1. Conceptual illustration of the Barker joint live-recapture/live-resight JLRLR model with capture events (gold dashed circles) and resight intervals (blue dashed circles). Animals (letters) are released with marks from the capture site (solid gold circle) during capture events and can disperse out of the capture site after initial release. The upper panel illustrates an ideal design, where the entire range is resightable whereas the lower panel illustrates a fixed site design where the capture site is excluded from resight. Note that after the release in the fixed site design, animal 'b' is highly resightable but not recapturable because of its location in space while animal 'a' is highly capturable but not resightable because of its location and animal 'e' is not resightable nor recapturable because of its location, thus illustrating a problematic sampling design.*

The JLRLR model assumes that resight probability is the same for all marked individuals, regardless of their location (Figure 1, top panel). In practice this means that resight probabilities should ideally be uniform across the possible range of all marked individuals, though it is unclear how violations to this assumption affect survival estimation. Other studies have used the JLRLR model to describe resights from public reports of tags over a prolonged interval (Barker 1997; Hall, McConnell & Barker 2001; Gibson *et al.* 2023), from agencies monitoring other geographic regions linked by dispersal/migration (Mizroch *et al.* 2004; Gibson *et al.* 2018; Hunt *et al.* 2018; Healy *et al.* 2022a), or technologies that continuously detect marked animals (Al-Chokhachy & Budy 2008; Bouwes *et al.* 2016; Beard *et al.* 2017; Gómez-Ramírez, Gutiérrez-González & López-González 2017). However, because resight efforts must typically vary spatially and may be clustered for many species, the assumption of spatially uniform resight probabilities is often violated. For example, resight probability can be influenced by observer heterogeneity (i.e., differences in resight probabilities due to spatial differences in observer bias, reporting probability, re-encounter probability, and migration rate), which has been identified as a major hurdle for studies of bird movement (Thorup & Conn 2009, Korner-Nievergelt *et al.* 2010). Additionally, the tendency for many animals to exhibit home range behaviors can influence detection and resight probabilities, as animals with home range centers close to detection points (e.g., trap locations) are more detectable (Royle & Young 2008).

Previous simulation work with the JLRLR model has found this model often outperforms CJS models for survival estimation and can provide reliable estimates of survival when resight data are continuous (Barbour, Ponciano & Lorenzen 2013) and animals emigrate from the capture site (Horton & Letcher 2008; Conner *et al.* 2015). Additionally, when dead recoveries are used in lieu of live resightings, heterogeneity in reporting rates of dead recoveries has been found to have minimal influence on survival bias (Nichols *et al.* 1982, Barker 1992). To our knowledge, however, no simulation studies have evaluated how violations to the uniform resight probability assumption for live resightings affects survival estimation (Barker, Burnham, & White 2004), particularly when fixed sites are used for resights and movement is non-random (e.g., movement around a home range or breeding site fidelity). Furthermore, we could find no guidance on how to best compile capture histories for the Barker JLRLR model in regards to visits to the capture site. Specifically, should visits to the capture site always be included as capture events (so that more marks could be included in the model), or was it necessary to have some visits to the capture site included as the resight interval to satisfy the assumption that marked animals are equally resightable? Accordingly, we use simulation to evaluate how sampling design choice (i.e., fixed versus random sites) and differences in resight probabilities inside and outside the capture site affect survival estimation in the Barker JLRLR model. We then introduce a new parameterization of a multistate Barker JLRLR model, where states correspond to sites inside and outside the sampling site and use the same simulated datasets to evaluate whether this approach improves survival bias.

We hypothesize that using fixed sites for resight can bias survival via two mechanisms. First, non-representation bias will occur when resight probabilities differ for individuals located inside and outside the capture site so that the resight probabilities for animals in the capture site are not representative of the population. Second, unobservable resight bias can occur when the area outside the capture site includes both resightable (observable) and non-resightable (unobservable) areas. If individuals cannot move between observable and unobservable resight areas, then survival will likely be underestimated as movement into the unobservable resight area is permanent and will be confounded with mortality (Kendall & Nichols 2002). If animals can move between observable and unobservable resight sites, then there is temporary emigration from the observable resightable areas, and this would most likely lead to terminal survival bias (i.e., survival bias that occurs at the end of the time series; Peñaloza, Kendall & Langtimm 2014).

Methods:

*Motivation of the simulation design*

Our simulations are motivated by our work in the Colorado River (Grand Canyon, Arizona, USA) where multiple agencies monitor fishes year-round using both fixed site (including sampling sites and sites with autonomous passive integrated transponder [PIT] antennas) and random site designs (Van

Haverbeke et al. 2017, Rogowski et al. 2018, Dzul et al. 2023). Adult survival estimates from models of humpback chub in a fixed site (capture site) in western Grand Canyon (hereafter wGC reach) were relatively low and it is suspected that the low estimate is due (at least in part) to emigration (Dzul et al. 2023). Information about fish marked in the wGC reach but resighted outside this reach by other monitoring efforts could be used in the Barker JLRLR model to help inform permanent emigration. However, even if resight information was utilized from other monitoring efforts (which mostly use fixed sites), there are large sections of river that are never sampled or sampled very infrequently. Our simulations are based on efforts to improve monitoring program design for estimating vital rates of fish populations using data from discrete field sampling efforts (i.e., sampling trips) and continuous detection data from autonomous antenna arrays (Beard *et al.* 2017; Pennock *et al.* 2020; Dzul *et al.* 2022).

*Single-state Barker joint live resight model*

We simulated capture histories in R (R Core Development Team 2020) and used Rmark (Laake 2013) to fit models in Program MARK (White & Burnham 1999) because MARK is commonly used by researchers (Barbour, Ponciano & Lorenzen 2013, Conner et al. 2015, Healy et al. 2022a) and allows for fast computation. Parameters in the single-state Barker model include $S$ (survival), $p$ (capture probability for animals in the capture site), $R$ (resight probabilities for live animals inside and outside the capture site), $F$ (probability of emigration from the capture site), $F'$ (probability of immigration into the capture site), and $r$ (probability a dead tag is reported). The type of emigration (i.e., temporary vs permanent) modeled is determined by the presence or lack of constraints on $F'$. Specifically, $F'$ is unconstrained to model temporary emigration, whereas $F'$ is fixed to zero to model permanent emigration. Additionally, to incorporate the continuous resight intervals into a discrete-time model structure, the JLRLR model introduces additional parameters ($R'$) for describing the probability an animal is resighted before dying, given it dies in the interval. The capture histories are paired so that first observation in each pair corresponds to a capture event at the capture site (1 if captured at the capture site, otherwise 0) and the second observation the resighting interval (2 if resighted between capture events, otherwise 0). For example, the encounter history '10 02' corresponds to an animal that was released in the capture site, not resighted during resight interval 1, not captured in the capture site during capture event 2, and resighted during resight interval 2. The likelihood for this encounter history would be: $(1 - R_1) \cdot S_1 \cdot (F_1 \cdot (1 - p_2) + (1 - F_1)) \cdot (S_2 \cdot R_2 + (1 - S_2) \cdot R'_2)$, because the animal survives between capture events 1 & 2 (as evidenced by being resighted after capture event 2), but it is unknown whether the animal is in the capture site and available for capture during capture event 2. Additionally, it is unknown whether or not the animal died during resight interval 2 and whether its resight probability is for animals that live (R) or die (R') during resight interval 2. Capture histories included 12 visits to the capture site (hereafter events) as well as 12 resight intervals for a total of 12 paired occasions (i.e., 24 time periods) where each occasion refers to a capture event followed by a resight interval. Note that occasion 1 corresponds to release-only (i.e., individuals could not be recaptured on the first occasion), but that during the following capture events, individuals released on previous events could be recaptured so that 11 capture probabilities and 12 resight probabilities are in the model structure.

Sampling sites used for simulations occurred along a river (i.e., 1-dimensional and ignoring which side of a river an individual is captured on and focusing on distances up or downriver along the center thalweg), but simulation results should be applicable to other terrestrial or aquatic landscapes. We simulated movement from a Cauchy distribution, a distribution that reflects the tendency for most individual movement to be short in distance with occasional long distance dispersal events, and which fits well observed movements in our well-sampled systems (Korman *et al.* 2016, Dzul, Yackulic, & Korman. 2018, Healy, Yackulic, & Schelly. 2022b). Animals could move among the 50 sites that occurred along a river. In the Barker JLRLR model, mortality and movement occur during the continuous resight interval so that animals can be resighted but die before the next capture event. To simulate this, we 'broke up' each capture-to-capture interval into two sub-intervals (capture-to-resight and resight-to-capture). This allowed animals to be resighted and die before the next capture event, thus imitating a continuous survival process across capture-to-capture intervals. Note that the survival probabilities estimated by the model correspond to the square of the simulated survival probabilities from sub-intervals (i.e., $S = 0.80$ over capture-to-resight and from resight-to-capture, corresponding to $S = 0.64$ from capture to capture).

During each capture event (i.e., odd-numbered time periods 1,3,5,…23), 100 animals were released in each of four sites (sites 21-24, hereafter the capture site). Between each capture event and resight interval, animals survived with probability 0.80 and could move to other sites based on the probabilities from a Cauchy distribution with location = 0 and scale parameter = 1, which is truncated so that animals cannot move out of the river[1]. If animals were in an observable resight site during the resight interval (see sampling scenario description for details), that animal could be resighted with probability 0.80, otherwise its resight probability was zero. This translates to a marginal resight probability that was typically 0.3-0.5. Between the resight interval and the next capture event, animals again survived and moved with the same probabilities as between capture and resight. Animals located in the capture site during capture events could be recaptured with probability 0.50. In addition to the above-described reference simulation set (ref), we also simulate the following scenarios: high survival (high-S) with $S = 0.9025$ from capture event to capture event, and high movement (high-move) where the scale parameter is set to 5 on the Cauchy distribution[2]. We simulated 100 data sets for each scenario.

To minimize biases associated with low capture and resight probabilities, we simulate using relatively high capture and resight probabilities as part of the main paper. These probabilities are not realistic for our study system, where capture probabilities are low and resights only occur in a small proportion of the river (i.e., low resight probability). To address this mismatch, we include Appendix A as alternative scenario (i.e., low observability/ low movement scenario) with lower movement, capture, and resight probabilities. To evaluate the effect of sample size, we run additional simulations where we decrease (25/site) or increase (400/site) the number of animals released in each of the four sites for the reference simulation set under permanent emigration (Appendix B).

---

[1] which translates to an emigration probability from the capture site of ~0.55-0.59 (conditional on survival) across capture events, and where ~49% of animals in the central site (i.e., can move no more than 25 units) move at least 1 spatial unit, ~10% move at least 5 spatial units, and ~4% move at least 10 spatial units.

[2] This translates to an emigration probability from the capture site of ~0.92-0.94 (conditional on survival) across capture events, and where ~85% of animals in the central site (i.e., can move no more than 25 units) move at least 1 spatial unit, ~43% move at least 5 spatial units, and ~19% move at least 10 spatial units.

*Resight sampling design*

We tested six different sampling designs (Figure 2, Appendix C), where scenarios differ in terms of which sites are selected for resightings: 1) global, 2) random, 3) fixed sites (including the capture site), 4) fixed sites (excluding the capture site), 5) random and fixed combinations (including the capture site as a fixed site), and 6) random and fixed combinations (excluding the capture site as a fixed site). Because most animals did not move out of the capture site, designs where resights could occur in the capture site (e.g., designs 3 & 5) inherently had a higher marginal resight probability than designs where resights could not occur in the capture site (i.e., designs 4 & 6). Thus, in an effort to make resight probabilities more comparable across designs, designs that included the capture site in resight had other resight sites located farther from the capture site, whereas designs that excluded the capture site had resight sites located in close proximity to the capture site. We provide a description (Appendix C) and an illustration (Figure 2) of the six different sampling designs evaluated. Note that designs with fixed and random sites (i.e., designs 5 & 6) have 26 resight sites compared to 24 resight sites for fixed-only designs (i.e., designs 3 & 4). The greater number of resight sites in fixed/random combo designs was to achieve a better balance of fixed vs random sites and to maximize spatial spread for fixed sites.

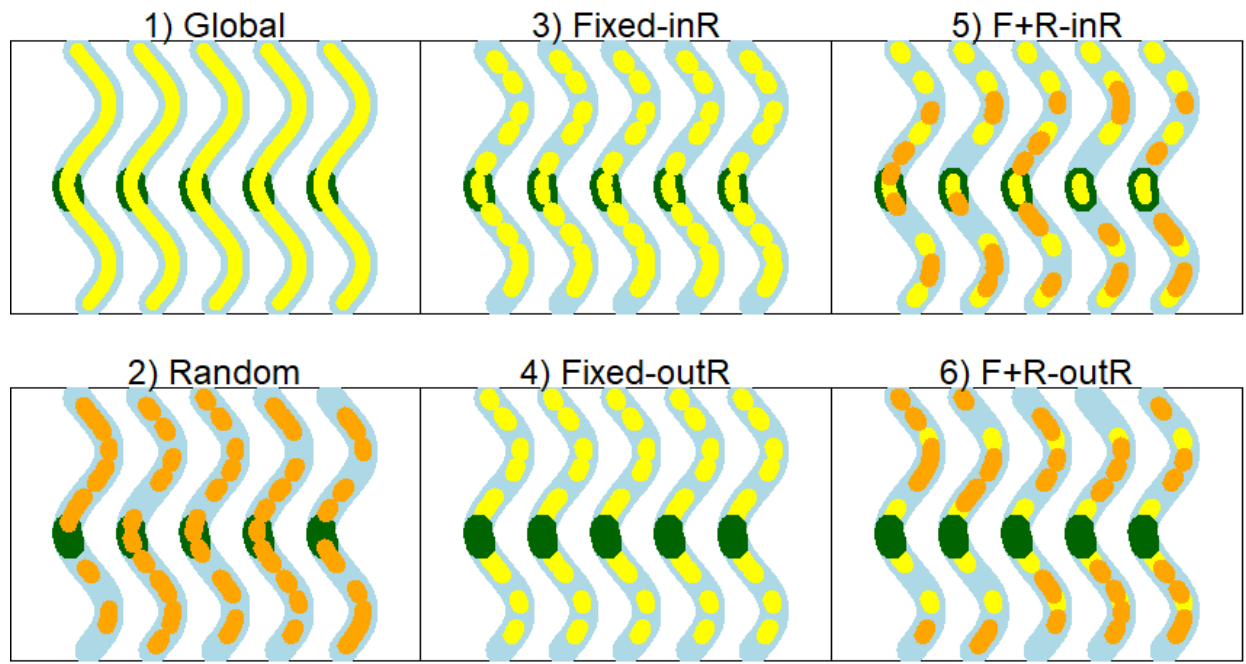

*Figure 2. Sampling design simulation scenarios for five hypothetical time periods, which are depicted as five wavy lines in each panel. The range of the study population is depicted as a river (blue) with a fixed capture site (green) and resight sites that are fixed (yellow) and(or) random (orange). The numbers refer to the scenario, as follows: 1) global resight, 2) random resight, 3) fixed resight that includes the capture site, 4) fixed resight that excludes the capture site, 5) fixed and random resight combination that includes the capture site, and 6) fixed and random resight combination that excludes the capture site.*

For the global design, resightings could occur uniformly throughout the entire river (i.e., in all 50 segments) with probability 0.32 (i.e., conditional resight probability (0.80) was multiplied by 0.40 of sites in resight). For the random design, 12 pairs of adjacent resight sites were randomly selected so that 24 sites were observable and 26 unobservable for each resight interval. Both fixed site designs (designs 3 & 4) also included 24 observable and 26 unobservable resight sites but differed in the location of sites used for resight and whether or not the capture sites were always included (design 3) or always excluded (design 4) from resight sites. Similarly, both sampling designs 5 & 6 included 26 sites for resightings (14 were fixed and 12 were randomly chosen), but differed in the site locations for fixed resight sites and whether or not the capture sites was included or excluded from resight (see Appendix C). In designs 5 & 6, random sites excluded sites that were already fixed sites.

The capture histories were modified to match the data type for the JLRLR model, so that all capture events included '1' if the individual was captured (otherwise 0), and all resight intervals included a '2' if the animal was resighted either inside or outside the capture site (otherwise 0). Because the fully time-dependent model with Markovian temporary emigration is not always estimable without additional movement constraints (Barker, Burnham & White 2004), the fitted JLRLR model included constant $S$ and time-specific estimates of $p$, $F$, $F'$, $R$, and $R'$, except that parameters for the last occasion were set equal to that of the second to last occasion to avoid parameter confounding. Because no dead recoveries were used in the model, we fixed $r = 0$. We fit two different model versions to the simulated data – 1) the permanent emigration model (where $F'$ is fixed to be 0) and animals therefore cannot re-enter the capture site, and 2) the immigration/emigration model, where animals can move back and forth into and out of the capture site. For both model types, parameter estimates are time-varying except that the last occasion is set equal to the second to last occasion to avoid confounding parameters.

We evaluate percent relative bias (i.e., $100 \times \left(\frac{true-estimated}{true}\right)$) in survival estimates from the Barker JLRLR model and compare it to apparent survival estimates from a CJS model that is fit to the same data but excludes resights. We also attempted to evaluate goodness of fit (GOF). To our knowledge, while there have been some GOF tests developed for the Barker model with dead recoveries (McCrea, Morgan, & Pradel 2014), there are no GOF tests specifically designed for the Barker JLRLR model with live resights. Because no contingency-table GOF tests were available, we evaluated GOF for simulated models using the median-ĉ simulation tool in Program MARK (White & Burnham 1999). According to Cooch (2008), this median-ĉ simulation tool is still a 'work in progress' that is useful for diagnosing lack of fit that is due to extra-binomial noise but may not be able to diagnose all different types of lack of fit. To determine whether this simulation tool could detect lack of fit in models with designs 3 & 4, we used 10 of the 100 data sets from each design for the reference scenario for both the permanent emigration and immigration/emigration models. The simulation tool creates capture histories with known ĉ and fits the saturated model to compare the deviances of the known- ĉ models to that of the observed model. For each dataset, five replicates were used at ĉ = 1.0, 1.5, 2.0, 2.5, and 3.0 to calculate mean ĉ, and we report the mean ĉ across the ten dataset replicates so that there are fifty replicates for each value of ĉ.

*Multi-state JLRLR model*

In attempt to minimize bias associated with non-representation bias, we developed a Bayesian multistate version of the Barker JLRLR model. While other parameterizations of the multistate Barker have been developed for Bayesian frameworks (Gibson *et al.* 2023) and in Program MARK (White & Burnham 1999), our multistate model allows for two transitions between occasions and includes spatially explicit resight probabilities that would minimize non-representation bias because $R$, $R'$, $S$, and movement (see below) probabilities can differ inside and outside the capture site. However, animals may still move between the observable and unobservable resight sites so that unobservable resight bias is still comparable to the single state version. The same simulated datasets used in the single state JLRLR model were reformatted to fit the multistate formulation (e.g., different states/indexing for animals resighted inside or outside the capture site).

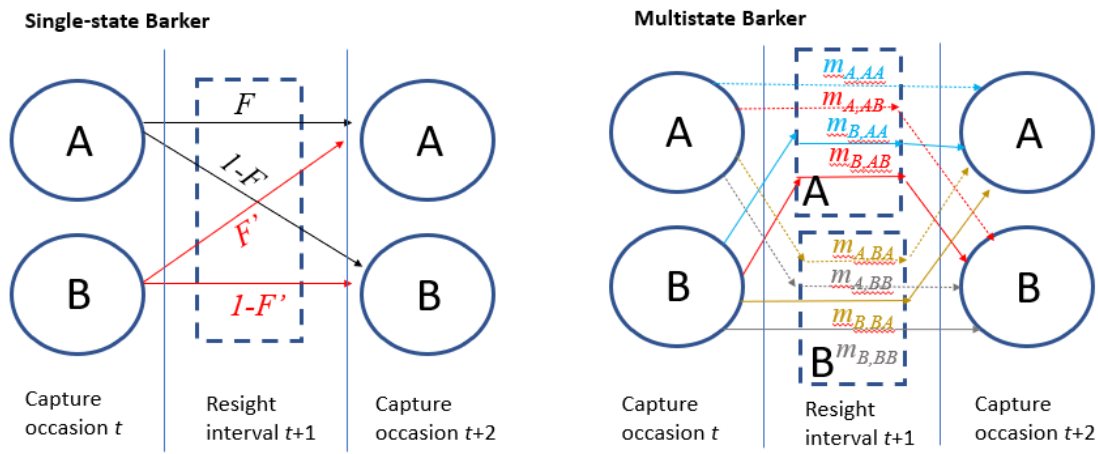

*Figure 3. Illustrative comparison of the movement parameterizations of the single-state and multistate Barker models, where survival is excluded (for simplicity) and A represents the area inside the capture site and B is outside the capture site. Whereas resight probabilities are equal for states A and B in the single-state model, the multistate version has resight as being different for states A and B and this requires a more complex movement model, where 'm' parameters describe movement that occurs between release, resight, and recapture. Note that $m_{A,BB} = 1-m_{A,AA}-m_{A,AB}-m_{A,BA}$ and $m_{B,BB} = 1-m_{B,AA}-m_{B,AB}-m_{B,BA}$.*

We constructed a multistate JLRLR model in Stan (Stan Core Development Team 2020) using the package rstan (Stan Development Team 2020), because the multistate version of the Barker model implemented in Program MARK differs from the formulation we are suggesting (i.e., MARK formulation only allows for one transition between capture events, whereas our model includes the possibility of two transitions). In our version, the two states correspond to whether animals are within (state A) or outside (state B) the capture site, with additional states for death (D). Animals could move between states A and B both between the capture event and resight interval, and again between resight interval and capture event, thus necessitating introducing secondary (2°) states (AA, AB, BA, BB, AD, BD, DD) where the first letter describes the state/location of the animal during the resight interval and the second letter describes the state/location during the following capture event. Animals released in state A at time *t* have $m_{A,AA}$ probability of being in 2° state AA, $m_{A,BA}$ probability of being in 2° state BA, $m_{A,AB}$ probability of being in state B, and $1- m_{A,AA} - m_{A,BA} - m_{A,AB}$ probability of being in 2° state BB (Figure 3). Movement probabilities for animals released in state B were estimated separately (i.e., $m_{B,AA}$, $m_{B,BA}$, $m_{B,AB}$, $1- m_{B,AA} - m_{B,BA} - m_{B,AB}$), to allow for applications where movement is influenced by memory. The *m* parameters were modeled using an m-logit link to ensure movement probabilities summed to 1. To describe mortality, recently dead animals could be resighted before death, so that animals released in A that died before the next capture event could be resighted in A with probability $(m_{A,AA} + m_{A,AB})R'_A$ and resighted in B with probability $(1-m_{A,AA} - m_{A,AB})R'_B$. We provide Stan code for this model in Appendix D. We used uniform priors for *p*, *R*, *R'*, and *S* and normal priors (μ = 0, σ = 1) for *m*. We ran models for 3 chains with 1000 iterations each (including 500 burn-in) for a total of 1500 posterior draws. Convergence was achieved with the Gelman Rubin statistic ($\hat{R}$) < 1.1.

*Results*

Single-state Barker JLRLR model

Simulation results for the permanent emigration model showed that the global and random study designs had minimal survival bias (mean percent relative bias ≤1%), though these designs exhibited substantial positive biases under low observability (Appendix A). Designs with fixed sites were either positively or negatively biased, depending on whether or not the capture site was included as part of the resight. Survival biases for the JLRLR model varied by scenario and were always stronger for the reference scenario compared to high-S and high-move (Figure 4, Figure 5). When the capture site was included in the resight efforts (i.e., designs 3 & 5), survival estimates were negatively biased. When the capture site was excluded from resighting (i.e., designs 4 & 6), survival estimates were positively biased for the reference scenario and unbiased or minimally biased for high-S and high-move. Sample size (i.e., the number of fish released) did not affect biases in survival estimates, though designs with lower sample sizes had poorer precision (Appendix B).

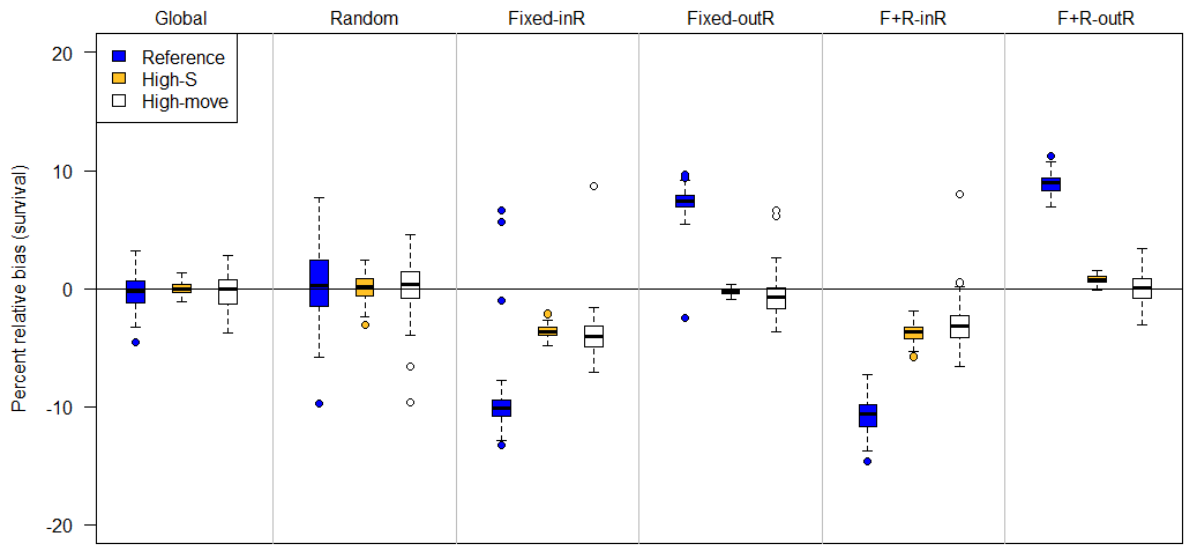

*Figure 4. Percent relative bias of the Barker joint live-recapture/live-resight model under six different sampling designs (top labels) and three scenarios (colors) when emigration is assumed to be permanent (F' = 0). The six sampling designs differed in the site selection for resight data: Global (entire area resightable), Random, Fixed-InR (fixed sites for resight that included the capture site), Fixed-OutR (fixed sites for resight that excluded the capture site), F+R-inR (fixed and random resight sites where the capture site was a fixed resight site), F+R-outR (fixed and random resight sites where the capture site was excluded from the resight). Compared to the reference (blue) scenario, high-S has higher survival (0.90 instead of 0.64), and high-move has more movement (scale on Cauchy parameter is 5 instead of 1). The upper and lower edges of each box represent interquartile range (IQR) and the bold line in the middle of each box is the median. The lower whiskers extend to which ever value is higher: minimum or the 25% quantile minus 1.5 times the IQR; whereas the upper whiskers extend to which ever value is lower: the maximum or the 75% quantile plus 1.5 times the IQR.*

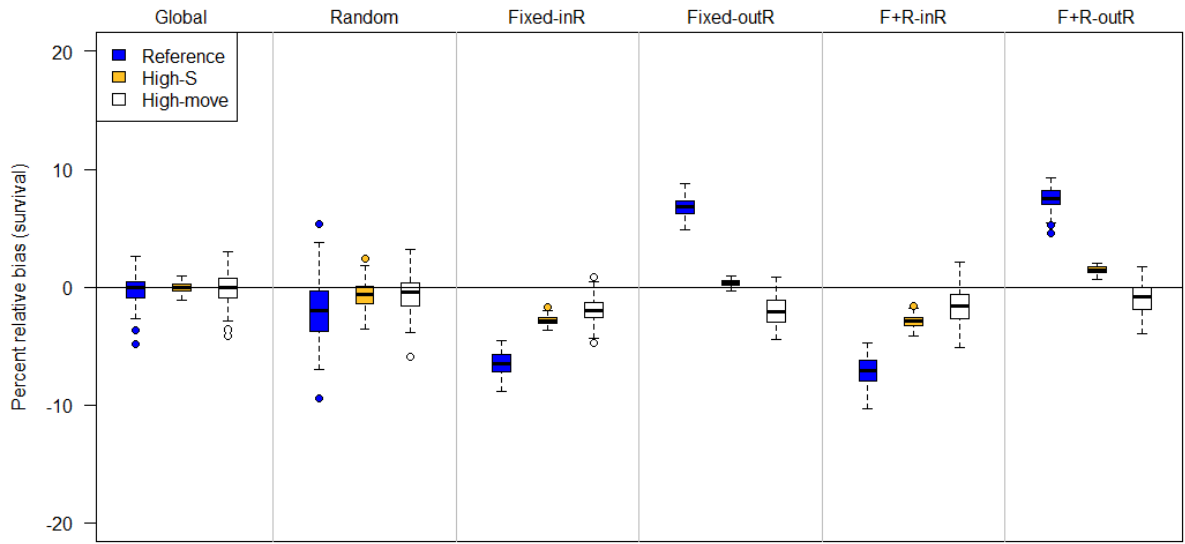

*Figure 5. Percent relative bias of the Barker joint live-recapture/live-resight model under six different sampling designs (top labels) and three scenarios (colors) when emigration and immigration are both estimated as part of the fitted model. The six sampling designs differed in the site selection for resight data: Global (entire area resightable), Random, Fixed-InR (fixed sites for resight that included the capture site), Fixed-OutR (fixed sites for resight that excluded the capture site), F+R-inR (fixed and random resight sites where the capture site was a fixed resight site), F+R-outR (fixed and random resight sites where the capture site was excluded from the resight). Compared to the reference (blue) scenario, high-S has higher survival (0.90 instead of 0.64), and high-move has more movement (scale on Cauchy parameter is 5 instead of 1). Some percent relative biases were -100% and are not apparent on the plot. The upper and lower edges of each box represent interquartile range (IQR) and the bold line in the middle of each box is the median. The lower whiskers extend to which ever value is higher: minimum or the 25% quantile minus 1.5 times the IQR; whereas the upper whiskers extend to which ever value is lower: the maximum or the 75% quantile plus 1.5 times the IQR.*

Simulation results from the immigration and emigration model (i.e., $F' \neq 0$) were generally similar to that of the permanent emigration model, except that the random design did exhibit minimal to moderate negative bias (-1 to -4%). Because results were relatively similar for the permanent emigration and immigration/emigration model, we report results from the permanent emigration model below and round bias to the nearest percentage. For design 3, mean percent relative bias in survival was -10% for the reference, -4% for high-S, and -4% for high-move scenarios, and for design 5 it was -11% for reference, --4% for high-S, and -3% for high-move (Figure 4). There was also positive bias in resight probability (Figure 6). Both designs that used fixed sites and excluded the capture site from resighting (designs 4 & 6) were also similar to each other. Mean survival biases for design 4 were +7% (reference), -1% (high-S), and 0% (high-move), and for design 6, mean survival biases were +9% (reference), +1% (high-S), and 0% (high-move). Design 4 also exhibited a slight positive bias in resight probability and a negative bias in capture probability (Figure 7).

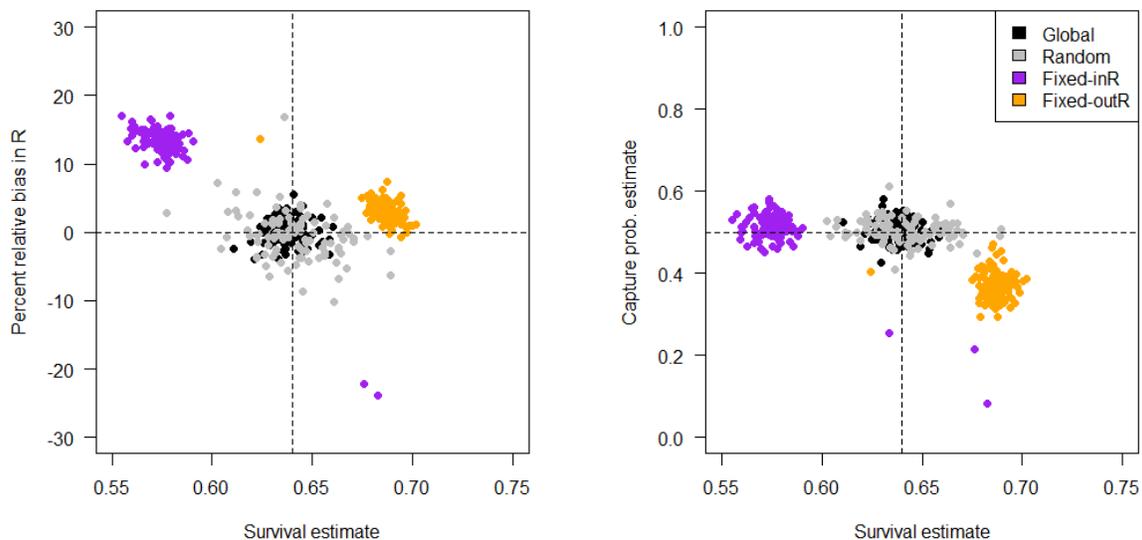

*Figure 6. Bias in resight (R) and capture probability (p) in relation to survival. Left panel: Percent relative bias (estimated value minus the true value divided by the true value of R multiplied by 100)*

*against survival estimates for sampling designs that differed in site selection for resight sites. Estimates are from a model with permanent emigration. The vertical dotted line represents the true survival probability used in simulation (0.64). Right panel: Capture probabilities (p) plotted against survival estimates for sampling designs that differed in site selection for resight sites. Estimates are from a model with permanent emigration.*

Although many Barker JLRLR models exhibited bias, survival bias from the JLRLR model was always less in magnitude than bias from a CJS model fit to the same datasets. For the CJS model, mean apparent survival estimates were 28% (reference), 39% (high-S), and 5% (high-move) for the permanent emigration model, which translates to percent relative biases of -65%, -66%, -59%, and -94%, respectively. For the immigration/emigration model, apparent survival estimates from the CJS model were 48% (reference), 72% (high-S), and 55% (high-move) with percent relative biases of -40%, -24%, and -31%, respectively. The CJS estimates of apparent survival also pertain to the datasets used for the multistate Barker simulation (see below).

For all replicates in each simulation, the ĉ simulation tool produced estimates of ĉ between 1.0-1.5, suggesting that the model fit the data adequately (Table 1). The highest ĉ occurred for design 4 (fixed never in resight). Graphs of deviance residuals produced by Program MARK were generally asymmetric, suggesting potential for lack of model fit.

| Model | Design | Scenario | Mean ĉ |
|---|---|---|---|
| Immigration/emigration | Fixed in resight | Reference | 1.24 |
| Immigration/emigration | Fixed never in resight | Reference | 1.33 |
| Permanent emigration | Fixed in resight | Reference | 1.27 |
| Permanent emigration | Fixed never in resight | Reference | 1.40 |

*Table 1. Mean ĉ values (across five replicates) estimated using the median ĉ simulation tool in Program MARK. In all ĉ tests, resight sites were fixed but differed in that they either included (fixed in resight, sample design 3) or excluded (fixed never in resight, sample design 4) the capture site. We repeated simulation tests for models with moderate capture probability (reference, p = 0.5) and low capture probability (low p, p = 0.2).*

Multistate Barker JLRLR model

Compared to the single-state JLRLR model, the multistate JLRLR model demonstrated less survival bias for most sampling designs (Figure 8, Figure 9). Bias was substantially reduced for the permanent emigration model for fixed site designs (designs 3 & 4), and the directionality of bias estimates was always negative. Specifically, the permanent emigration model displayed negligible to moderate bias for the global design (0% reference, -1% high-S, -3% high-move), slightly positive to moderately negative when the capture site was included in resight (1% for reference, -1% for high-S, and -5% for high-move), and low to moderate negative bias when the capture site was excluded from resight (-3% for reference, -2% for high-S, and -6% for high-move). The immigration/emigration model also showed reduced bias for fixed site designs, but interestingly had negative bias for the global design (-1% for reference, 0% for high-S, and -3% for high move). Also, the bias pattern for fixed sites in design 3 where the capture site was included in resight were either unbiased or slightly biased (+1% for reference, 0% for high-S, and -1% for high-move). When the capture site was excluded from resight (design 4), biases were all negative (-3 for reference, -2% for high-S, and -1% for high-move).

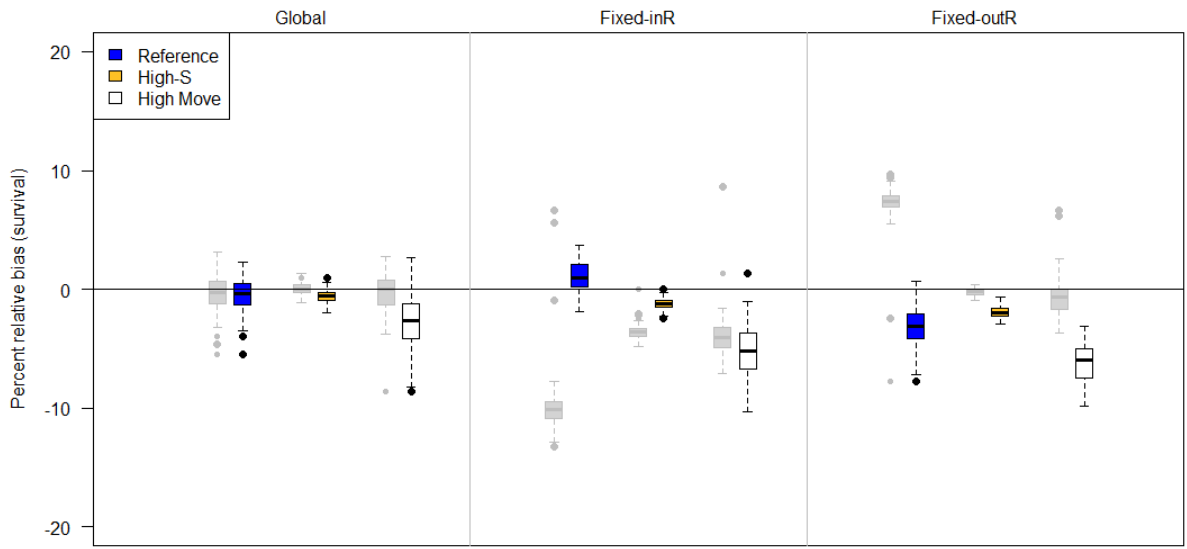

*Figure 8. Percent relative bias of the multistate Barker joint live-recapture/live-resight (JLRLR) model with permanent emigration under three different sampling designs (top labels) and three scenarios (colors). The adjacent grey boxplots describe survival estimates from the corresponding single state Barker JLRLR model with an assumption of permanent emigration. The three sampling designs differed in the site selection for resight data: Global (entire area resightable), Fixed-InR (fixed sites for resight that included the capture site), and Fixed-OutR (fixed sites for resight that excluded the capture site). Compared to the reference (blue) scenario, high-S has higher survival (0.90 instead of 0.64), and high-move has more movement (scale on Cauchy parameter is 5 instead of 1). The upper and lower edges of each box represent interquartile range (IQR) and the bold line in the middle of each box is the median. The lower whiskers extend to which ever value is higher: minimum or the 25% quantile minus 1.5 times the IQR. Whereas the upper whiskers extend to which ever value is lower: the maximum or the 75% quantile plus 1.5 times the IQR.*

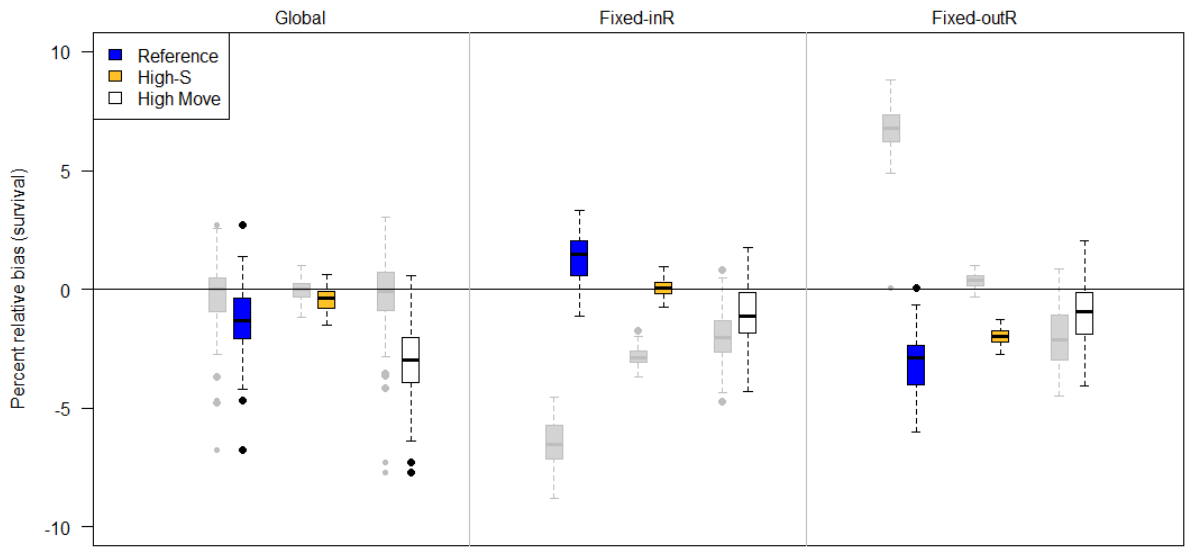

*Figure 9. Percent relative bias of the multistate Barker joint live-recapture/live-resight (JLRLR) model under three different sampling designs (top labels) and three scenarios (colors) when emigration and immigration are both estimated. The adjacent grey boxplots describe survival estimates from the corresponding single state Barker JLRLR model with immigration and emigration estimated. The three sampling designs differed in the site selection for resight data: Global (entire area resightable), Fixed-InR (fixed sites for resight that included the capture site), and Fixed-OutR (fixed sites for resight that excluded the capture site). Compared to the reference (blue) scenario, high-S has higher survival (0.90 instead of 0.64) and high-move has more movement (scale on Cauchy parameter is 5 instead of 1). The upper and lower edges of each box represent interquartile range (IQR) and the bold line in the middle of each box is the median. The lower whiskers extend to which ever value is higher: minimum or the 25% quantile minus 1.5 times the IQR. Whereas the upper whiskers extend to which ever value is lower: the maximum or the 75% quantile plus 1.5 times the IQR.*

Discussion:

Our simulation results demonstrate the potential for both non-representation bias and unobservable resight bias in survival estimation when resight probabilities are heterogeneous due to non-random animal movement and use of fixed sites for resight. Furthermore, even when fixed sites are not used for resight, low resight and capture probabilities can bias survival (Appendix A). Most previous simulation work evaluating individual heterogeneity has focused on closed population models or open models with discrete capture occasions and no auxiliary resight data (e.g., CJS models). With these models, individual heterogeneity in capture probabilities is a well-known source of negative bias in abundance estimation (Edwards & Eberhardt 1967; Otis *et al.* 1978), whereas negative survival bias, though present, is minimal/negligible (Carothers 1979; Pledger, Pollock & Norris 2010). Perhaps for these reasons, modelling of individual heterogeneity is often considered when inferences are focused on abundance (Chao 1988; McClintock, White & Burnham 2006) but is less common in open models of survival (but see Pledger et al. 2010).

Our results suggest that using a single-site Barker JLRLR model can lead to substantial non-representation bias in survival when animals in the capture site have resight probabilities that are non-representative of the population. Specifically, when the capture site was included in the resighting effort, captures were positively linked with resights (i.e., animals located in the capture site were more likely to be captured and resighted), and survival was underestimated. When the capture site was excluded from resight, the capture process was negatively related to the resight process (i.e., animals in the capture site were more capturable but less resightable than outside) and survival was overestimated. By using a multistate Barker JLRLR model with different resight probabilities inside and outside the capture site, the capture and resight processes were spatially separated so that resight probabilities in the study site could differ from those outside the study site and this minimized non-representation bias.

Furthermore, our results suggest that unobservable resight bias was also present in Barker JLRLR models, as evidenced by negative survival biases when the capture site was included in resight (designs 3 & 5), and the observation that these designs led to a subset of animals that became unobservable (i.e., animals that emigrated from the capture site to a site where no resight occurred). This type of bias is similar to that produced by CJS, where emigration and survival are confounded (i.e., apparent survival). However, note that survival estimates from designs 3 & 5 in the Barker JLRLR model were more accurate than survival estimates from CJS models, suggesting that Barker JLRLR survival estimates could serve as a lower bound for true survival that could account for some (but not all) permanent emigration from the capture site.

Horton and Letcher (2008) used simulation to compare survival estimates from Cormack-Jolly-Seber (CJS), Barker JLRLR, and robust design models for different movement patterns and found that the JLRLR model estimates proved less biased for scenarios with high emigration. However, they assumed all resight probabilities were constant (no spatial variability) and resight probabilities were simulated to be very high (0.75). In practice, it is rare to obtain resight probabilities as high as 0.75 and to ensure resight efforts occur throughout the entire range. Conner et al. (2015) also used simulation to compare the bias of the CJS and Barker JLRLR models under various movement scenarios but assumed either that resight probabilities were spatially uniform or removed known emigrants (after they emigrated) from the analysis. Unlike Horton and Letcher (2008), Conner et al (2015) did find bias in survival estimates from the Barker JLRLR model but concluded that the JLRLR model performed as good or better than the CJS model.

While it could be argued that the Barker JLRLR model is an improvement over CJS (as evidenced by the higher magnitude of bias in the latter), there are several factors that should be considered when comparing these two model types. First, the CJS model produces estimates of apparent survival, thus acknowledging that emigration and survival are confounded (Lebreton *et al.* 1992). In contrast, survival estimates from the JLRLR model are frequently interpreted as true survival (Barker 1997, Cooch 2008), though our simulation results suggest this only occurs under global or random resight designs. Second, whereas the directionality of bias is known for the CJS model (i.e., negative bias), our simulation results indicate that bias for the JLRLR model may be either positive or negative. This positive survival bias can mask survival declines by leading biologists to falsely conclude that low apparent survival probabilities are due to high permanent emigration from study area and not high mortality. The multistate JLRLR model formulation described in this paper may be a good alternative to the single state version when fixed sites are used for resight and animal movement is non-random. However, compared to the single-state version, the multistate model includes more nuisance parameters associated with movement and resight and, while our simulations illustrated that it reduced bias compared to the single state version, both positive and negative biases were still present. Another alternative for species with one-time ontogenetic movement dispersal (e.g., natal dispersal in birds) is the mark-recapture approach with natal dispersal described by Badia-Boher et al. (2023).

Despite being undesirable from a resight perspective, fixed sites can be the most convenient, cost-effective strategy for monitoring, particularly for stationary detection technologies (e.g., PIT antennas, Motus towers [Taylor *et al.* 2017], camera traps). In light of our simulation results, we advise that if fixed sites are used for resight, the capture site should be included in resight (to avoid positive survival bias) and survival estimates should be reported as estimates of apparent survival. Our simulation results also illustrate that there may be some conditions when fixed site resight is a suitable sampling design, for example with animals that exhibit higher mobility and(or) random movements so that animal behavior induces more mixing (i.e., randomness) in the fixed site design. Models of animals with moderate movement and high survival may also produce less biased survival estimates when they include many occasions. Situations where all animals leave the capture site (e.g., neotropical migrant birds) may also be better suited for fixed site resight, though more simulation work may be required to address how Markovian movement between breeding/overwintering habitat affects survival estimation.

Detections from long, continuous intervals are becoming increasingly common with the growing popularity of citizen science projects (Metcalfe *et al.* 2022; Swinnen *et al.* 2022) and remote detection technologies (Evans *et al.* 2020; Dzul *et al.* 2021; Gilbert *et al.* 2021). Our study illustrates that although these types of auxiliary resightings have the potential to improve population models (Van Strien, Van Swaay & Termaat 2013; Robinson *et al.* 2018), study design and modeling assumptions must be evaluated to avoid or anticipate biased parameter estimates. New data analysis tools, such as continuous-time mark-recapture models (Fouchet *et al.* 2016) and data integrated models (Besbeas *et al.* 2002; Schaub & Abadi 2011), may be well-suited for handling these types of data but likewise require considerations of how violation of model assumptions inherent in the sampling process may impact inferences.

*Acknowledgements*

We thank two reviewers for their helpful comments on an earlier draft version. The Glen Canyon Dam Adaptive Management Program and the Bureau of Reclamation helped fund this study. Any use of trade, firm, or product names is for descriptive purposes only and does not imply endorsement by the U.S. Government.

Appendix A. Simulation assessment of survival estimates under low observability and movement

Methods

We refer to this simulation set as 'low observability' (or LO) based on the lower value for capture and resight probabilities used in simulations, and we compare these to the 'high observability' (or HO) set described in the main paper. Specifically, for the LO set, emigration was permanent, capture probabilities were 20%, conditional resight probabilities were 20% in resight sites (0% outside resight sites), and 16% of the river was visited during resight (Figure A1, Table A1). This translates to mean marginal resight probabilities between 3.2-3.3% (design 1: global), 3.9-5.1% (design 2: random), 8.8-17.3% (design 3: fixed in resight), 1.3-3.4% (design 4: fixed outside resight) for all three scenarios. Note that in the LO set, the fixed in resight design had higher resight probabilities, in part due to animals being less mobile and therefore more likely to remain in the capture site.

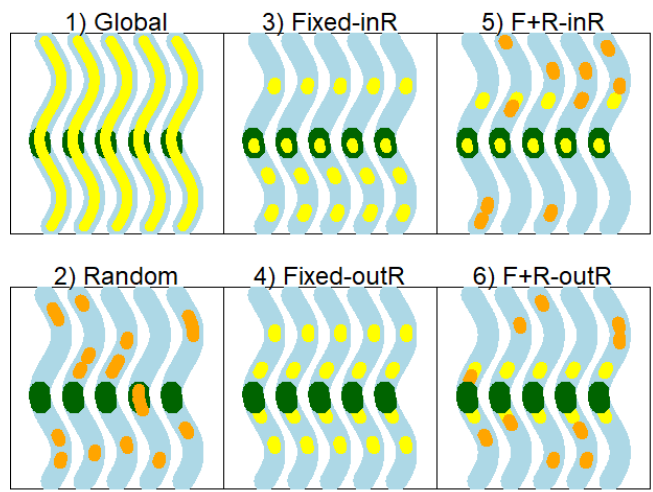

*Figure A1. Sampling design simulation scenarios for low observability for five hypothetical time periods, depicted as five wavy lines in each panel. The range of the study population is depicted as a river (blue) with a fixed capture site (green) and resight sites that are fixed (yellow) and(or) random (orange). The numbers refer to the scenario, as follows: 1) global resight, 2) random resight, 3) fixed resight that includes the capture site, 4) fixed resight that excludes the capture site, 5) fixed and random resight combination that includes the capture site, and 6) fixed and random resight combination that excludes the capture site.*

| Design | Name | Number of fixed sites | Number of random sites | Fixed sites | Random sites* |
|---|---|---|---|---|---|
| 1 | Global | 50 | 0 | 1-50 | - |
| 2 | Random | 0 | 8 | - | 1-50 |
| 3 | fixed sites (including CS) | 10 | 0 | 3&4, 13&14, 21&22&23&24, 37&38 | - |
| 4 | fixed sites (excluding CS) | 8 | 0 | 9&10, 19&20, 29&30, 39&40 | - |
| 5 | random and fixed combination (including CS as a fixed site) | 6 | 4 | 21&22&23&24, 33&34 | 1-20, 25-32, 35-50 |
| 6 | random and fixed combination (excluding CS as a fixed site) | 4 | 4 | 19&20, 29&30 | 1-18, 25-28, 31-50 |

*Table A1. Description of different sampling designs used for simulations. Sampling designs differ by sites selection strategy (either global, fixed, random, or fixed/random combination) for resightings and whether or not the capture site (CS) was included or excluded as a resight site.*
*\*Note that for random sites, only a subset of sites (see column 'Number of random sites') are actually visited.*

In addition to resight and capture probabilities, the scale of the movement parameter in the Cauchy distribution is also substantially lower (scale = 0.2 instead of 1) for the LO simulation set and corresponds to an emigration probability of ~9-10% across capture events. The lower movement probability was chosen in effort to more closely resemble movement of humpback chub, *Gila cypha*, which are relatively sedentary (Kaeding et al. 1990, Paukert et al. 2006, Gerig et al. 2014) and have low resight probabilities outside the capture site. However, we retain one scenario in the LO set ('LO: high-move') that has movement equal to that used in the reference scenario of the high observability set ('HO: ref'), thus allowing for comparison of results based on observability only between the two simulation sets.

LO simulations were similar to the HO set in terms of the number of releases and survival probabilities. Specifically, during each capture event (i.e., odd-numbered time periods 1,3,5,…23), 100 animals were released in each of four sites (sites 21-24, hereafter the capture site) and survival was 64% between capture events. Between the resight interval and the next capture event, animals again survived and moved with the same probabilities as between capture and resight. In addition to the above-described reference simulation set (LO: ref), we also simulate the following scenarios: high survival (LO: high-S) with survival = 90.25%, and high movement (LO: high-move) where the scale parameter is set to 1 on the Cauchy distribution.

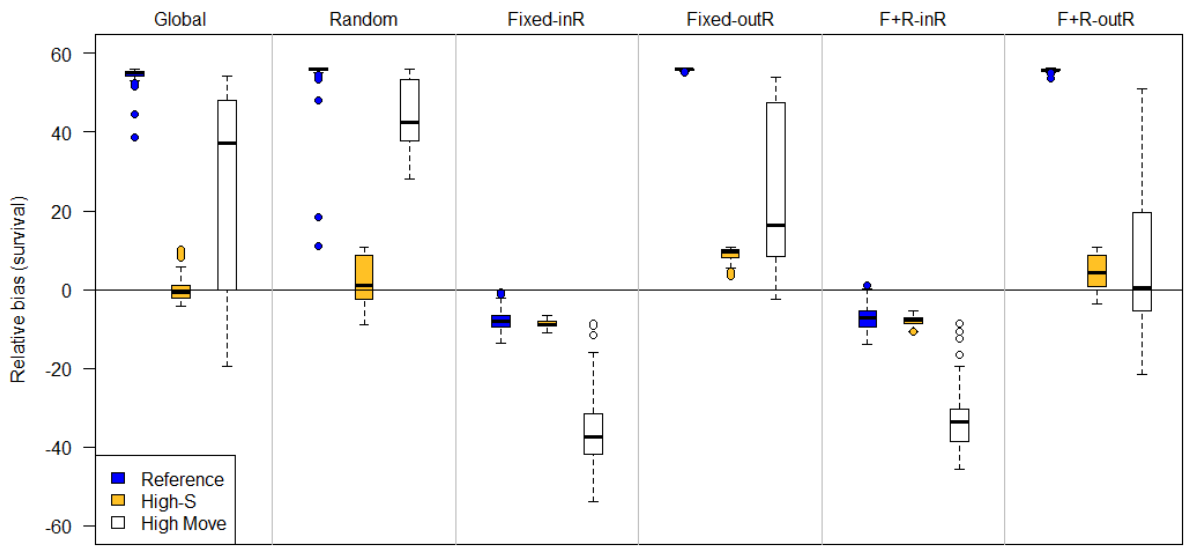

*Figure A2. Percent relative bias of the Barker joint live-recapture/live-resight model under six different sampling designs (top labels) and four scenarios (colors) when emigration is permanent (F' = 0) and observability is low. The six sampling designs differed in the site selection for resight data: Global (entire area resightable), Random, Fixed-InR (fixed sites for resight that included the capture site), Fixed-OutR (fixed sites for resight that excluded the capture site), F+R-inR (fixed and random resight sites where the capture site was a fixed resight site), F+R-outR (fixed and random resight sites where the capture site was excluded from the resight). Compared to the reference (blue) scenario, high-S has higher survival (90% instead of 64%), and high-move has more movement (scale on Cauchy parameter is 1 instead of 0.2). The upper and lower edges of each box represent interquartile range (IQR) and the bold line in the middle of each box is the median. The lower whiskers extend to which ever value is higher: minimum or the 25% quantile minus 1.5 times the IQR; whereas the upper whiskers extend to which ever value is lower: the maximum or the 75% quantile plus 1.5 times the IQR.*

Results:

The LO: high-S scenario generally produced the least biased estimates compared to LO: ref and LO: high-move, so we focus mostly on discussion of LO: ref and LO: high-move. Unlike the HO set, which found that random and global designs produced negligble bias, these two designs produced positive bias in survival estimates in the LO simulations, where the mean percent relative bias was 54.6% and 54.9% for LO: ref for global and random designs, respectively (Figure A2). The LO: high-move scenario also exhibited high percent relative bias for global and random designs (25.9% and 44.6%, respectively). Additionally, the LO set produced more severe positive biases for designs where the fixed resight (design 4) and fixed-random combo (design 6) excluded the capture site. Specifically, percent relative bias for LO-ref was 55.9% (design-4) and 55.7% (design 6) and for LO-high move it was 25.0% (design 4) and 7.8% (design 6). In contrast, the fixed resight and fixed-random combo designs that include the capture site (i.e., designs 3 & 5) displayed negative survival bias, with -7.9% (design 3) and -7.3% (design 5) for LO-ref and -36.2% (design 3) and -33.5% (design 5) for LO: high-move. For comparison, percent relative bias in survival for a CJS model for LO: ref was -9.6% and for LO: high-move was -56.5%, suggesting a slight to moderate improvement in bias with the Barker JLRLR model for designs where fixed resight sites included the capture site.

Discussion:

Results from the LO set provides additional insight into survival estimation with the Barker JLRLR model. Notably, even relatively 'good' designs (i.e., global & random) can display large biases when resight and capture probabilities are low. While biases are less apparent when survival is high, survival is often unknown and certain designs (e.g., designs 1, 2, 4, 6) can over-estimate survival, so that justifying the design based on the survival estimate may lead to error. The designs with fixed sites that included the capture site (designs 3 & 5) inherently had higher resight values and displayed bias patterns that were similar to the HO set, where survival was negatively biased. To avoid positive survival biases under low observability, the capture site should be included as part of resight (i.e., design 3) and estimates of survival should be reported as apparent survival.

Appendix B. Effect of sample size on bias in survival

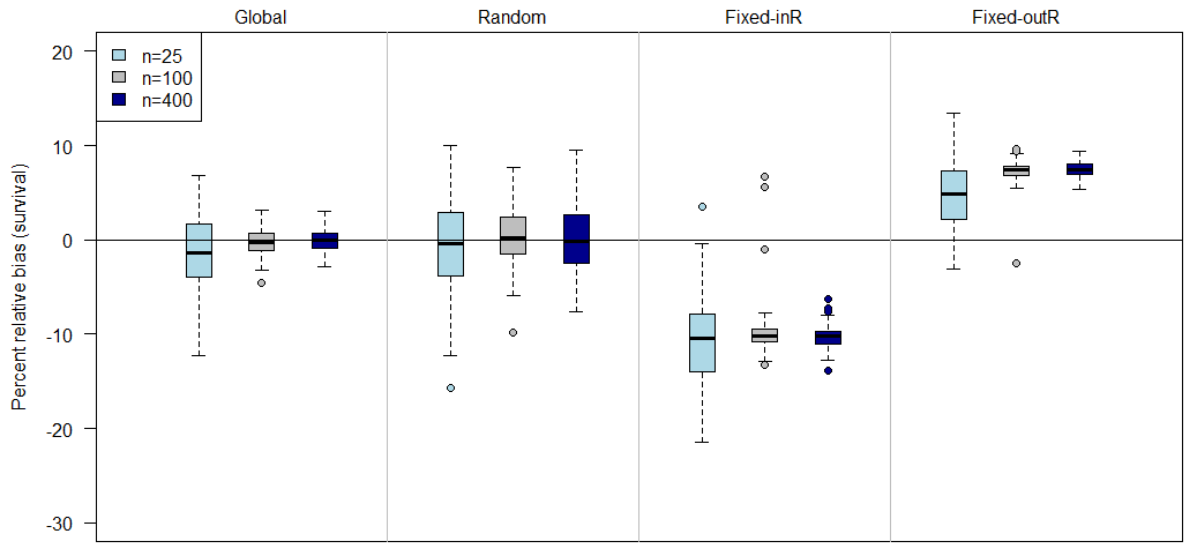

*Figure B1. Relative bias in survival estimates from the Barker joint live resight live recapture model under different sampling designs (top labels) and with different numbers of animals released in each site (n). Sampling designs describe the sites where animals could be resighted and include Global (all animals have same resight probability, regardless of location), Random (sites are randomly chosen throughout all potential locations), Fixed-InR (fixed sites are chosen and include the capture site), and Fixed-outR (fixed sites are chosen and exclude the capture site).*

We evaluated the effect of sample size (i.e., number of animals marked and released) on survival estimation in the Barker joint live resight live recapture model (JLRLR). Simulations were identical to the reference scenario (with permanent emigration) described in the associated manuscript, where survival was equal to 64% (from capture to capture occasion), capture probability was 50%, and resight probability was 80%, the scale parameter for the Cauchy distribution was set to 1, and 100 animals were released in each site on each occasion. To evaluate the effect of sample size, we ran scenarios with lower numbers of animals released (25 per site per occasion) and higher numbers of animals released (400 per site per occasion).

Results showed no noticeable difference between the reference (n=100) and high sample size (n=400) scenarios. The low sample scenario (n=25) showed greater variance in mean survival estimates across simulations. This result is expected because lower numbers of tagged animals leads to lower numbers of recaptures and higher uncertainty in survival. The directionality of the bias for the low sample size scenario was comparable to the reference and high sample size scenarios, where bias was negligble for global and random designs, negative for designs with fixed resight sites that included the capture site, and positive for designs with fixed resight sites that excluded the capture site.

Appendix C. Sites used for resight in different sampling designs

| Design | Name | Number of fixed sites | Number of random sites | Fixed sites | Random sites* |
|---|---|---|---|---|---|
| 1 | global | 50 | 0 | 1-50 | - |
| 2 | random | 0 | 24 | - | 1-50 |
| 3 | fixed sites (including CS) | 24 | 0 | 3&4, 6&7, 9&10, 13&14, 17&18, 21&22&23&24, 27&28, 33&34, 37&38, 43&44, 47&48 | - |
| 4 | fixed sites (excluding CS) | 24 | 0 | 3&4, 9&10, 15&16, 17&18, 19&20, 25&26, 27&28, 29&30, 35&36, 39&40, 45&46, 49&50 | - |
| 5 | random and fixed combination (including CS as a fixed site) | 14 | 12 | 1&2, 11&12, 21&22, 23&24, 33&34, 43&44, 49&50 | 3-10, 13-20, 25-32, 35-42, 45-48 |
| 6 | random and fixed combination (excluding CS as a fixed site) | 14 | 12 | 9&10, 17&18, 19&20, 25&26, 27&28, 41&42 | 1-8, 11-16, 29-40, 43-50 |

*Table C1. Description of different sampling designs used for simulations. Sampling designs differ by sites selection strategy (either global, fixed, random, or fixed/random combination) for resightings and whether or not the capture site (CS) was included or excluded as a resight site.*
*\*Note that for random sites, only a subset of sites (see column 'Number of random sites') are actually visited.*